\documentclass{aa}
\usepackage[varg]{txfonts}
\usepackage{natbib}
\usepackage{graphicx}
\usepackage{enumitem}
\usepackage{xcolor}
\usepackage{setspace}
\usepackage{amsmath,bm}
\usepackage{url}
\newcommand{\angstrom}{\text{\normalfont\AA}}
\bibpunct{(}{)}{;}{a}{}{,} % to follow the A&A style
\begin{document}

\title{A polarimetric study of asteroids in comet-like orbits}  
        \author{Jooyeon Geem\inst{1,2}
                \and Masateru Ishiguro\inst{1,2}
                \and Yoonsoo P. Bach \inst{1,2}
                \and Daisuke Kuroda\inst{3}
                \and Hiroyuki Naito\inst{4}
                \and Hidekazu Hanayama\inst{5}
                \and Yoonyoung Kim\inst{6}
                \and Yuna G. Kwon\inst{6}
                \and Sunho Jin\inst{1,2}
                \and Tomohiko Sekiguchi\inst{7} 
                \and Ryo Okazaki\inst{7}
                \and Jeremie J. Vaubaillon\inst{8}
                \and Masataka Imai\inst{9}
                \and Tatsuharu Oono\inst{10}
                \and Yuki Futamura\inst{10}
                \and Seiko Takagi\inst{10}
                \and Mitsuteru Sato\inst{10}
                \and Kiyoshi Kuramoto\inst{10}
                \and Makoto Watanabe\inst{11}}

\institute{Department of Physics and Astronomy, Seoul National University, 1 Gwanak, Seoul 08826, Republic of Korea
            \and SNU Astronomy Research Center, Seoul National University, 1 Gwanak-ro, Gwanak-gu, Seoul 08826, Republic of Korea \\ \email{\textcolor{black}{geem@astro.snu.ac.kr, ishiguro@astro.snu.ac.kr}}
                \and Okayama Observatory, Kyoto University, 3037-5 Honjo, Kamogata, Asakuchi, Okayama 719-0232, Japan
                \and Nayoro Observatory, 157-1 Nisshin, Nayoro, Hokkaido 096-0066, Japan
                \and Ishigakijima Astronomical Observatory, National Astronomical Observatory of Japan, 1024-1 Arakawa, Ishigaki, Okinawa 907-0024, Japan
                \and Institut f\"{u}r Geophysik und Extraterrestrische Physik, Technische
Universit\"{a}t Braunschweig, Mendelssohnstr. 3, 38106 Braunschweig, Germany
                \and Asahikawa Campus, Hokkaido University of Education, Hokumon, Asahikawa, Hokkaido 070-8621, Japan
                \and Observatoire de Paris, I.M.C.C.E., Denfert Rochereau, Bat. A., FR-75014 Paris, France
                \and Faculty of Science, Kyoto Sangyo University, Banyukan B401, Motoyama, Kamigamo, Kita-Ku, Kyoto-shi, Kyoto 603-8555, Japan
                \and Department of Cosmosciences, Graduate School of Science, Hokkaido University, Kita-ku, Sapporo, Hokkaido 060-0810, Japan
                \and Department of Applied Physics, Okayama University of Science, 1-1 Ridai-cho, Kita-ku, Okayama, Okayama 700-0005, Japan}

        \date{Received 13 August, 2021. / Accepted XX October, 2021.}

\abstract{Asteroids in comet-like orbits (ACOs) consist of asteroids and dormant comets. Due to their similar appearance, it is challenging to distinguish dormant comets from ACOs via general telescopic observations. Surveys for discriminating dormant comets from the ACO population have been conducted via spectroscopy or optical and mid-infrared photometry. However, they have not been conducted through polarimetry.} {We conducted the first polarimetric research of ACOs.}
    {We conducted a linear polarimetric pilot survey for three ACOs: (944) Hidalgo, (3552) Don Quixote, and (331471) 1984 QY1. These objects are unambiguously classified into ACOs in terms of their orbital elements (i.e., the Tisserand parameters with respect to Jupiter  $T_\mathrm{J}$ significantly less than 3). Three ACOs were observed by the 1.6 m Pirka Telescope from UT 2016 May 25 to UT 2019 July 22 (13 nights).}
    {We found that  Don Quixote and Hidalgo have polarimetric properties similar to comet nuclei and D-type asteroids (optical analogs of comet nuclei). However, 1984 QY1 exhibited a polarimetric property consistent with S-type asteroids. We conducted a backward orbital integration to determine the origin of 1984 QY1, and found that this object was transported from the main belt into the current comet-like orbit via the 3:1 mean motion resonance with Jupiter. }
    {We conclude that the origins of ACOs can be more reliably identified by adding polarimetric data to the color and spectral information. This study would be valuable for investigating how the ice-bearing small bodies distribute in the inner Solar System.}

\maketitle

\section{Introduction}  
        A classification between comets and asteroids (the notation is given in Appendix \ref{app:Usage of term}) is important for investigating the compositional distribution in the present Solar System. In a conventional view, asteroids are distributed in the inner Solar System (i.e., mostly located in the main-belt region with low eccentricities), while comets originate from the outer Solar System (the Kuiper Belt or the Oort Cloud) with high eccentricities.

Because of their different origins, asteroids and comets have been conventionally distinguished by several properties. In terms of appearance, comets show tails and comae by ejecting gas and dust as they approach the Sun. Asteroids generally do not show cometary activity (except active asteroids, \citealt{Jewitt_2012}), so they have a point-source appearance. In terms of the orbital properties, the Tisserand parameter (an approximation derived from the Jacobi integral of the circular restricted three-body problem) with respect to Jupiter ($ T_\mathrm{J}$) has been employed to discriminate between comets and asteroids. In general, asteroids are dynamically disconnected from Jupiter, while comets are coupled or intersect with the orbit of Jupiter, providing $ T_\mathrm{J} > 3 $ for asteroids and $ T_\mathrm{J} < 3 $ for comets \citep{1982BAICz..33..104K,1997Icar..127...13L}.
In terms of the optical properties, reflectance spectra and geometric albedos ($p_\mathrm{V}$) are used for classification \citep{2008A&A...487.1195L,Licandro2011,2014ApJ...789..151K,2008Icar..194..436D}. Typically, comet nuclei have red spectra with low $p_\mathrm{V}$ ($ p_\mathrm{V} = 0.02$--$0.06 $, \citealt{2002EM&P...89..117C,2004come.book..223L}), and asteroids have a wide range of reflectance spectra and geometric albedos ($ p_\mathrm{V} =  0.02$--$0.60$,  \citealt{Usui2011}). However, it turns out that this conventional classification could not work for some objects, such as asteroids in comet-like orbits (i.e., asteroids having apparent $ T_\mathrm{J} < 3 $, hereafter ACOs).

Although dormant comets in the ACO population have been investigated via several methods, such as optical multiband photometry, spectroscopy, and infrared photometry \citep{fernandez2001, 2005AJ....130..308F,2014ApJ...789..151K}, few have been studied through polarimetry. Polarimetric observations  can provide the polarization degree--phase angle $P_\mathrm{r}$$(\alpha$) profile of targets, where $\alpha$ is the Sun--target--observer angle. In general, the $P_\mathrm{r}(\alpha)$ profiles of small bodies in the Solar System show negative $P_\mathrm{r}$ (i.e., light polarized in a parallel direction to the scattering plane) at $\alpha\lesssim\alpha_0$ and positive $P_\mathrm{r}$ (i.e., the perpendicularly polarized direction with respect to the scattering plane) at $\alpha \ge \alpha_{0}$ \citep{2016MNRAS.455.2091C}. Here, $ \alpha_0 $ is the inversion angle where $P_\mathrm{r} (\alpha_{0})= 0$ is established, and which generally appears at $\alpha \sim 20\degr$. Then, $P_\mathrm{r}$ pseudolinearly increases around $\alpha_{0}$ with a slope of $h$ and shows the maximum polarization degree ($P_\mathrm{max}$) at $\alpha_\mathrm{max}\sim100\degr$. Consequently, the $P_\mathrm{r}(\alpha)$ profiles are characterized by several key parameters (e.g., the slope $h$, $\alpha_0$, $P_\mathrm{max}$). The surface properties (such as albedo and grain size) were conjectured with these parameters. \citep{1986MNRAS.218...75G,Dollfus1989, 1992Icar...99..468S, Lupishko2018}.

In this paper we conducted a polarimetric pilot survey of three ACOs, (944) Hidalgo, (3552) Don Quixote, and (331471) 1984 QY1 (hereafter Hidalgo, Don Quixote, and QY1), to test the potential of polarimetry for ACO research. We chose Don Quixote and QY1 not only because they were bright in 2016--2019, but also because they have a high probability of being Jupiter-family comets \citep[$>96$ $\rm{\%}$,][]{2002Icar..156..399B}. Hidalgo is a dormant comet candidate because of its orbital and spectral properties \citep{Hartmann1987, Tholen1984}. More detailed information on the targets is summarized in Table 1.
We describe the observations and data reduction processes in Sect. \ref{observation} and observational results in Sect. \ref{sec:results}. We discuss the results based on our polarimetry and the dynamical properties and the surface prospects of the polarimetric study for ACOs in Sect. \ref{sec:discussion}.

\begin{table*}
                \caption{Orbits and spectral types of our targets}
                \label{T:list}
                \centering
                \begin{tabular}{l c c c c c c}
                        \hline\hline
                        Target Name & $a^a$ & $e^b$ & $i^c$ & $T_\mathrm{J}^d$ &  Spectral type & References\\
                                &(\mbox{au})& &(\mbox{deg})&  &  &  \\
                        \hline
                        QY1 & 2.50 & 0.89 & 14.3 & 2.68 & Unidentified (S$_\mathrm{q}$ or Q)$^e$ & \ldots \\
                         Don Quixote & 4.26 & 0.71 & 31.1 & 2.31 & D & 1, 2\\
                        Hidalgo & 5.73 & 0.66 & 42.5 & 2.07 & D & 2 \\
                        \hline
                \end{tabular}
                \tablefoot{                                  
                        \tablefoottext{a}{Semimajor axis,}
                        \tablefoottext{b}{Eccentricity,}
                        \tablefoottext{c}{Inclination,}
                        \tablefoottext{d}{Tisserand parameter with respect to Jupiter,}
                        \tablefoottext{e}{See Sect. \ref{sec:qy1}.}
                        We obtained these elements ($a$, $e$, and $i$) from the web-based JPL Small-Body Database Browser (https://ssd.jpl.nasa.gov/sbdb.cgi).}
                \tablebib{(1) \citet{Binzel2004}; (2) \citet{1989aste.conf..298T}.
                %(1)~\citet{Bus2002};
}
        \end{table*}

\section{Observations and data analysis}
        \label{observation}

\begin{figure}
                \resizebox{\hsize}{!}{\includegraphics{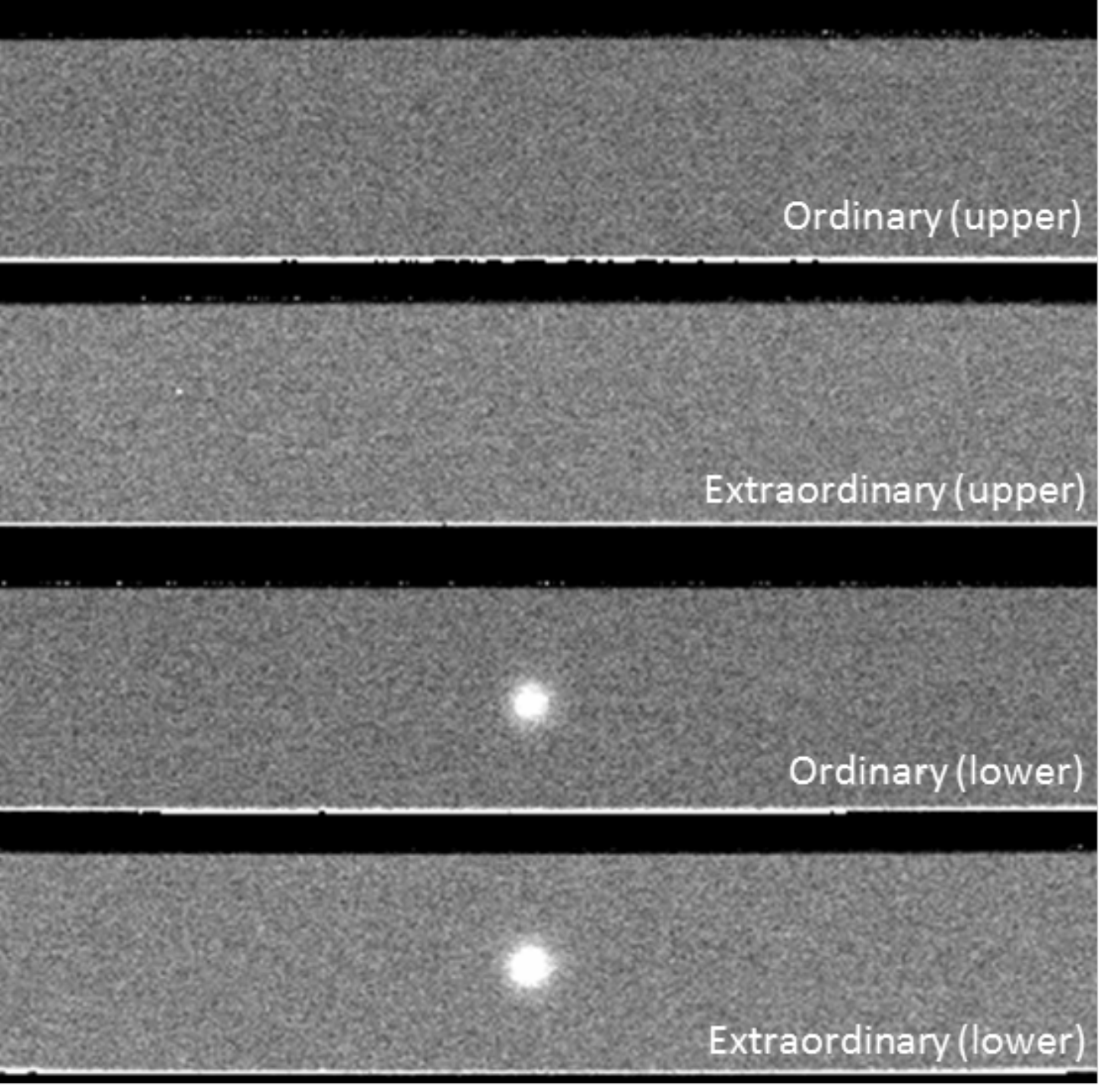}}
                \caption{ Example of a preprocessed polarimetric image of QY1 taken in the $ R_\mathrm{C} $-band with an exposure time of 120 $\mbox{sec}$. The field of view is split into two regions (upper and lower) by the polarization mask (see Sect. \ref{observation}).}
                \label{figure 1}
        \end{figure}

The journal of the observations is given in Table 2. We conducted polarimetry with the 1.6-$\mbox{m}$ Pirka Telescope at the Hokkaido University Observatory (142\fdg5 E, 44\fdg4 N at 151 $\mbox{m}$ above sea level, observatory code number Q33) in Japan from UT 2016 May 25 to 2019 July 22. During this period there were three ACOs (Don Quixote, Hidalgo, and QY1) that were bright enough to be measured by the instruments ($V$-band magnitudes $\lesssim$17 $\mbox{mag}$) with sufficiently small errors ($\lesssim$ 1 $\rm{\%}$) at moderately large $\alpha \sim 30\degr$. Among them, we had an opportunity to observe QY1 at a very large $\alpha$ ($\alpha\approx100\degr $). We utilized the visible multispectral imager (MSI) attached to the Cassegrain focus of the 1.6-m Pirka telescope, covering a field of view of $3\farcm3 \times 3\farcm3$ with a pixel resolution of $0\farcs39$ \citep{2012SPIE.8446E..2OW}. We obtained the polarimetric data with $V$-band and $R_\mathrm{C} $-band filters. The polarimetric module is optional for MSI, which consists of a rotatable half-wave plate and a Wollaston prism (a polarizing beam splitter), which has the advantage of reducing the influence of time-dependent atmospheric extinction. A polarization mask divides the field of view into two areas of the sky of $3\farcm3$ $\times$0\farcm7, and each area produces a data set having ordinary and extraordinary images simultaneously without mixing ordinary and extraordinary signals (Fig. \ref{figure 1}). We adjusted an exposure time of 60--180 $\mbox{sec}$, considering the signal-to-noise ratio at each half-wave plate angle (changed in the sequence of $ \theta= 0\degr, 45\degr, 22\fdg5$, and $67\fdg5$).

\begin{table*}
\caption{Observation circumstance}
\label{table:1}
\centering
\begin{tabular}{l c c c c c c c c c}
\hline\hline
Target Name & Date& UT & Filter & Exptime$^a$ & $N^b$ & $ r^c $& $ \Delta^d $&$ \alpha^e $& $ \phi^f $\\
&& & &($\mbox{sec}$) & &($ \mbox{au}$)&($ \mbox{au}$)&(\mbox{deg})&(\mbox{deg})\\
\hline
QY1 & 2016-May-25 & 12:09--17:20 & $ R_\mathrm{C} $ &120--180 & 28 & 0.87 & 0 30 & 111.7 & 42.2\\
& 2016-May-27 & 13:22--14:40 & $ V $ & 120 & 20 & 0.90 & 0.28 & 104.8 & 60.6\\
& 2016-May-27 & 12:53--14:19 & $ R_\mathrm{C} $ & 120 & 24 & 0.90 & 0.28 & 104.8 & 60.5\\
& 2016-May-28 & 12:55--15:42 & $ V $ & 120 & 28 & 0.92 & 0.28 & 101.0 & 74.0 \\
& 2016-May-28 & 13:35--16:26 & $ R_\mathrm{C} $ & 120 & 28 & 0.92 & 0.28 & 101.0 & 74.5 \\
& 2016-May-29 & 12:58--16:29  & $ V $ & 120 & 28 & 0.94 & 0.28 & 97.1 & 89.1 \\
& 2016-May-29 & 12:50--16:16 & $ R_\mathrm{C} $ & 120 & 20 & 0.94 & 0.28 & 97.1 & 89.3 \\
& 2016-Jun-21 & 12:54--14:38 & $ V $ & 60 & 60 & 1.33 & 0.52 &43.4 &131.3 \\
& 2016-Jun-21 & 11:25--17:24 & $ R_\mathrm{C} $ & 60--120 & 56 & 1.33 & 0.52 & 43.4 &131.3 \\
& 2016-Jun-24 & 11:20--14:56 & $ R_\mathrm{C} $ & 60--120 & 32 & 1.38 & 0.58 & 41.2 & 128.8 \\
\hline
Don Quixote & 2018-Jul-24 & 17:18--18:09 & $ R_\mathrm{C} $ & 120--180 & 20 & 1.58 & 1.46 &38.7 &254.0\\
& 2018-Aug-28 & 15:57--16:16 & $ R_\mathrm{C} $ & 60 & 20 & 1.86 & 1.40 &32.5 &248.4\\
& 2018-Sep-01 & 14:53--18:18 & $ R_\mathrm{C} $ & 60 & 88 & 1.89 & 1.40 &31.5 &246.5\\
& 2018-Sep-02 & 15:57--17:41 & $ I_\mathrm{C} $ & 120 & 36 & 1.90 & 1.40 &31.2 &246.0\\
\hline
Hidalgo & 2018-Sep-01 & 17:13--17:59 & $ R_\mathrm{C} $ & 60 & 32 & 2.02 & 1.90 &29.6 &266.5\\
&2019-Apr-19 & 13:15--13:59 & $ R_\mathrm{C} $ & 120 & 24 & 3.52 & 4.49 &22.9 &119.0\\
&2019-May-30 & 12:13--13:05 & $ R_\mathrm{C} $ & 120 & 24 & 2.82 & 2.92 &20.3 &105.4\\
&2019-Jul-22 & 11:18--11:43 &  $ R_\mathrm{C} $ & 120 & 12 & 10.93 & 3.79 &13.1 &93.9\\
\hline

\end{tabular}
  \tablefoot{
\tablefoottext{a}{Exposure time for each image in $\mbox{sec}$,}
\tablefoottext{b}{Number of exposures used to obtain polarimetric parameters,}
\tablefoottext{c}{Median heliocentric distance in $ \mbox{au}$,}
\tablefoottext{d}{Median geocentric distance in $ \mbox{au} $,}
\tablefoottext{e}{Median solar phase angle in \mbox{deg},}
\tablefoottext{f}{Position angle of the scattering plane in \mbox{deg}.}\\
The web-based JPL Horizon system (http://ssd.jpl.nasa.gov/?horizons) was used to obtain these quantities.}
 \end{table*}

 \begin{table*}
\caption{Polarimetric results}
\label{table:2}
\centering
\begin{tabular}{l c c c c c c c c c c c}
\hline\hline
Target Name&Date& UT & Filter&$\alpha$ &$ P^a$ & ${\sigma P}^b$ & $ {\theta_\mathrm{P}}^c $ & ${\sigma\theta_\mathrm{P}}^d $& ${P_\mathrm{r}}^e $ & ${\theta_\mathrm{r}}^f $\\
&&&&(\mbox{deg})&($ \% $) &($ \% $)&(\mbox{deg})&(\mbox{deg}) &($ \% $)&(\mbox{deg})\\
\hline
QY1 & 2016-May-25 & 12:09--17:20 & $ R_\mathrm{C} $ & 111.7 & 8.48 & 1.08 & $-50.95$ &3.64 & 8.45 & $-2.63$ \\
& 2016-May-27 & 13:22--14:40 & $ V $ & 104.8  & 8.00 & 0.31 & $-31.94$ & 1.12 & 7.97 & $-2.63$ \\
& 2016-May-27 & 12:53--14:19 & $ R_\mathrm{C} $ & 104.8 & 7.97 & 0.21 & $-32.76$ & 0.75 & 7.92 & $-3.33$ \\

& 2016-May-28 & 12:55--15:42 & $ V $ & 101.0  & 8.56 & 0.33 & $-19.56$ & 1.09& 8.56 & $-4.14$  \\
& 2016-May-28 & 13:35--16:26 & $ R_\mathrm{C} $ & 101.0  & 8.48 & 0.27 & $-17.20$ & 0.90 & 8.46 & $-1.88$  \\

& 2016-May-29 & 12:58--16:29 & $ V $ & 97.1  & 7.72 & 0.24 & $-4.74$ & 0.90 & 7.65 & $-3.85$ \\
& 2016-May-29 & 12:50--16:16 & $ R_\mathrm{C} $ & 97.1 & 7.64 & 0.27 & $-2.19$ & 0.99 & 7.63 & $-1.56$ \\

& 2016-Jun-21 & 12:54--14:38 & $ V $ & 43.4  & 2.88 & 0.35 & 42.45 & 3.48 & 2.88 & 1.14 \\
& 2016-Jun-21 & 11:25--17:24 & $ R_\mathrm{C} $ & 43.3 & 2.75 & 0.38 & 38.06 & 3.98 & 2.73 & -3.25 \\

& 2016-Jun-24 & 11:20--14:56 & $ R_\mathrm{C} $ & 41.2  & 2.32 & 0.55 & $32.27$ & 6.81 & 2.26 & -6.56 \\  
\hline
Don Quixote & 2018-Jul-24 & 17:18--18:09  & $ R_\mathrm{C} $ & 38.7 & 7.64 & $0.37$ & $-21.48$ & $1.39$& $7.50$& $-5.50$\\
& 2018-Aug-28 & 15:57--16:16 & $ R_\mathrm{C} $ & 32.5 & 4.65 & $0.71$ & $-30.40$ & $4.39$& $4.43$& $-8.80$\\
& 2018-Sep-01 & 14:53--18:18 & $ R_\mathrm{C} $ & 31.5 & 3.72 & $0.44$ & $-23.74$ & $3.41$& $3.72$& $-0.25$\\
& 2018-Sep-02 &  15:57--17:41 & $ I_\mathrm{C} $ & 31.2 & 4.16 & $0.63$ & $-20.19$ & $4.32$& $4.12$& $3.85$ \\
\hline
Hidalgo & 2018-Sep-01 & 17:13--17:59 & $ R_\mathrm{C} $ & 29.6 & 3.47 & $0.12$ & $-5.52$ & $0.97$& $3.46$& $-1.99$\\
& 2019-Apr-19 & 13:15--13:59 & $ R_\mathrm{C} $ & 22.9 & 1.16 & $0.14$ & $23.45$ & $3.49$& $1.14$& $-5.51$\\
& 2019-May-30 & 12:13--13:05 & $R_\mathrm{C}$ & 20.3 & 0 & $0.31$ & $-11.80$ & $49.49$& $0$& $-27.22$\\
&2019-Jul-22 & 11:18--11:43& $R_\mathrm{C}$ & 13.1 & 0.91 & $0.51$ & $-86.37$ & $16.03$& $-0.91$& $89.75$\\
\hline
\end{tabular}
\tablefoot{
\tablefoottext{a}{Nightly averaged linear polarization degree in percent,}
\tablefoottext{b}{Error of $P$ in percent,}
\tablefoottext{c}{Position angle of the strongest electric vector in \mbox{deg},}
\tablefoottext{d}{Error of $ \theta_\mathrm{P} $ in \mbox{deg},}
\tablefoottext{e}{Polarization degree referring to the scattering plane in percent (see Eq. (3)), }
\tablefoottext{f}{Position angle referring to the scattering plane in \mbox{deg} (see Eq. (4)).}
}
\end{table*}
        
We analyzed the data in the same manner as in \citet{Ishiguro17} and other papers using Pirka/MSI \citep{2015ApJ...814..156K, kuroda2021}. The raw data were bias-subtracted and flat-fielded using the MSI data reduction package. Cosmic rays were subtracted using the \texttt{L.A. Cosmic} tool \citep{2001PASP..113.1420V}. After preprocessing, we performed aperture photometry to extract the source fluxes from the ordinary and extraordinary parts of objects on the images using the photometry package in the Image Reduction and Analysis Facility (\texttt{IRAF}) and \texttt{astropy} \citep{2013A&A...558A..33A, Astropy2018} of the Python package. The typical aperture size was $2\farcs73$--$5\farcs85$. Additionally, by visual inspection, we excluded the images with background objects within an aperture radius from the center of our targets.

We obtained the linear polarization degree ($P$) and the position angle ($\theta_\mathrm{P}$) with the  equations
        \begin{equation}
        P = \sqrt{\left(\frac{Q}{I}^2\right) + \left(\frac{U}{I}^2\right)}~~
        \label{eq:P}
        \end{equation}
        and
        \begin{equation}
        \theta_\mathrm{P} = \frac{1}{2}\tan^{-1}\left(\frac{U}{Q}\right) ~~,
        \end{equation}
        where $I$, $ Q $, and $ U $ are the Stokes parameters derived from the extracted fluxes \citep{Tinbergen1996}, and $\theta_\mathrm{P}$ denotes the polarization position angle with respect to the celestial north. Before deriving $P$ and $\theta_\mathrm{P}$ in the above equations, we corrected the polarization efficiency, instrumental polarization, and position angle offset at the given wavelengths \citep{Ishiguro17}. After the calibration, we calculated the weighted mean of $q$ and $u$ on each date and derived the polarimetric result. To consider the influence of random noise in $P$, we applied the following equation \citep{Wardle1974}:
        \begin{equation}
        P' = \sqrt{P^2 - \sigma^{2}_\mathrm{P}}~~.
        \end{equation}
        When $P$ is nearly equal to zero, $\sigma^{2}_\mathrm{P}$ inevitably becomes larger than $P^2$, which makes the value in the root negative. In this case, we regard it as $P = 0 \%$.

Finally, we converted a polarization degree with respect to the scattering plane (the plane constituted by the target, the Sun, and the Earth) with the  equations
        \begin{equation}
        P_\mathrm{r} = P' \cos \left( 2\theta_\mathrm{r} \right) ~~ 
        \end{equation}
        and
        \begin{equation}
        \theta_\mathrm{r} = \theta_\mathrm{P} - \left( \phi \pm 90\degr \right) ~~,
        \end{equation}

\noindent where $ \phi $ is the position angle referring to the scattering plane on the sky and  $\theta_\mathrm{r}$ is the angle between the measured direction of the strongest electric vector and the normal to the Sun--target--observer plane, following the convention of asteroids and comet polarimetry (e.g., \citealt{Lupishko2014}). The $ \pm $ sign in parentheses is chosen to satisfy $
0\degr \le (\phi \pm 90\degr) \le 180\degr $ \citep{1993Icar..103..144C}.

All the preprocessed polarimetric data used are available via CDS\footnote{\url{http://cdsweb.u-strasbg.fr/cgi-bin/qcat?J/A+A/XXX}}.

\section{Results}
\label{sec:results}
\begin{figure*}
                \centering
                \resizebox{6cm}{!}{\includegraphics{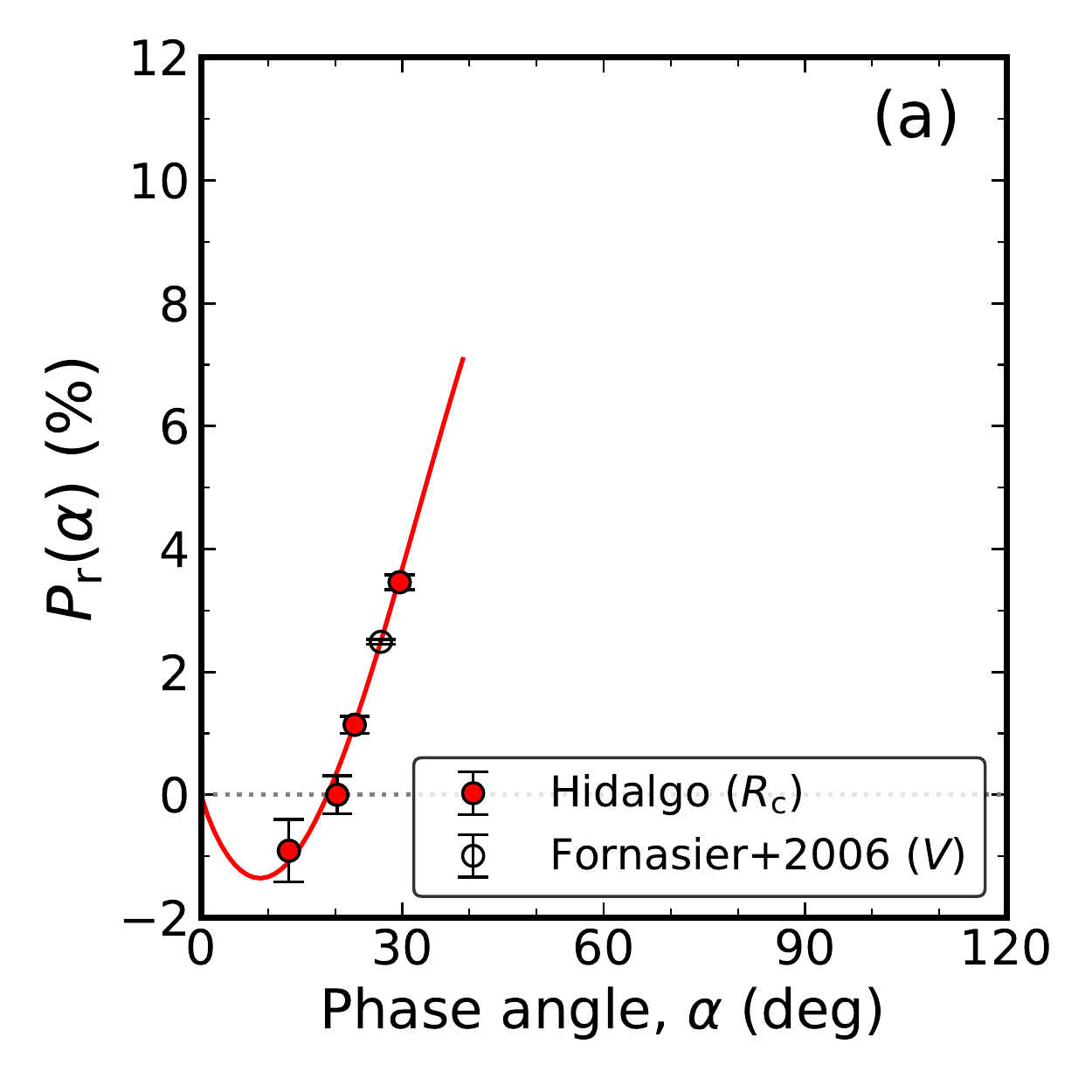}}
                \resizebox{6cm}{!}{\includegraphics{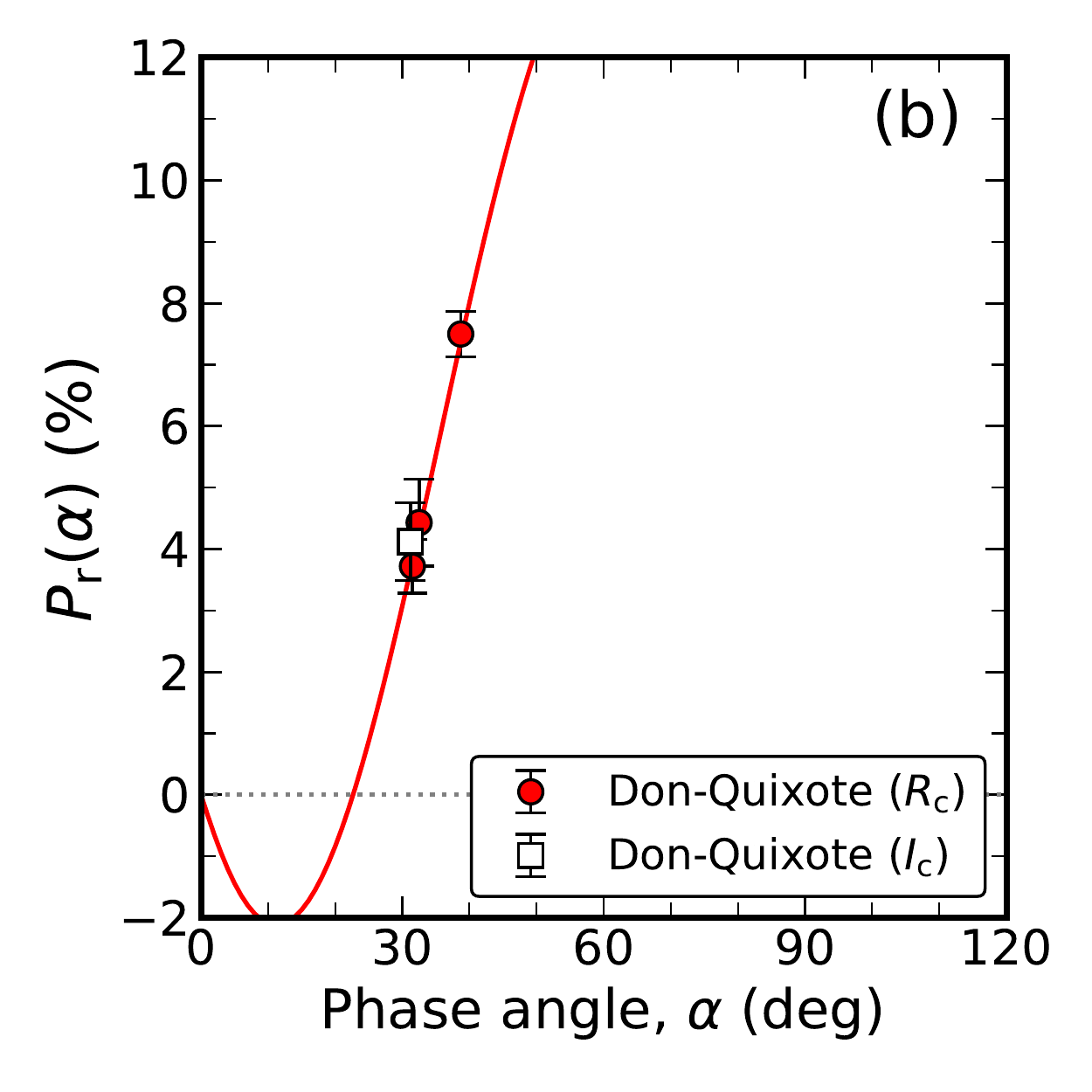}}
                \resizebox{6cm}{!}{\includegraphics{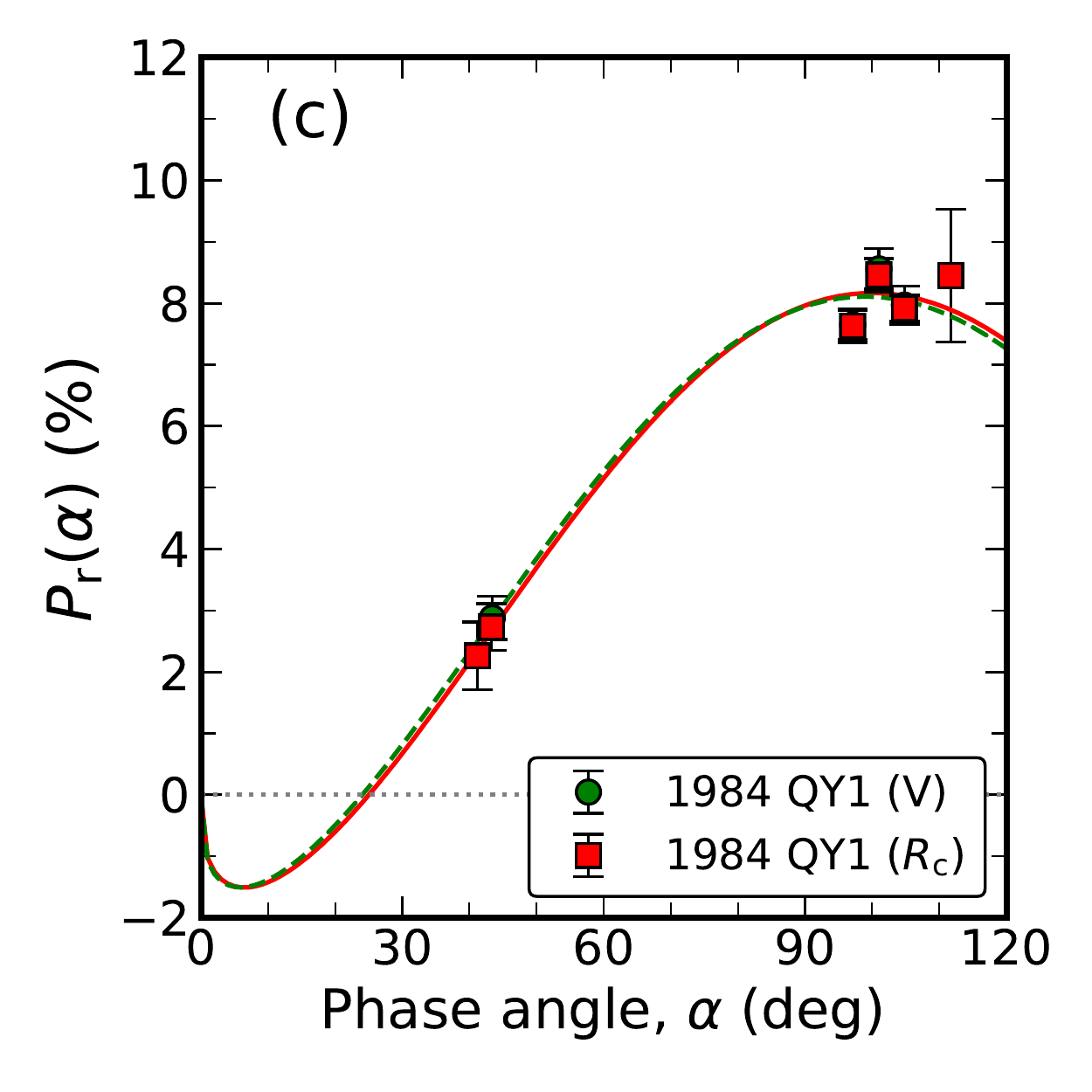}}
                \caption{Phase angle dependence of nightly averaged $P_\mathrm{r}$ for three ACOs: (a) Hidalgo, (b) Don Quixote, and (c) QY1. Polarization phase curves using the median of the Monte Carlo samples are shown as the red solid line for $R_\mathrm{C} $ band and the green dashed line for $V$ band by using Eq. (\ref{eq:trig}). For Hidalgo  we plot the $V$-band data (open circle) from \citet{Fornasier2006}.}

                \label{Figure 2}
        \end{figure*}

Table \ref{table:2} summarizes the weighted mean values of the nightly linear polarimetric result and Fig. \ref{Figure 2} shows the polarization phase curve. In the following subsections we describe our findings.

%Fitting the Pr profile
\subsection{Characterization of polarization phase curves}
\label{Pr-alpha pro}

To capture the outlines of polarization phase curves, we fit the data using the empirical  Lumme--Muinonen function \citep{Lumme1993,2005A&A...432.1081P}. It is given as
        \begin{equation}
        \label{eq:trig}
        P_\mathrm{r} (\alpha)= h \left( \frac{\sin \left( \alpha \right)}{\sin \left( \alpha_{0} \right)}\right)^{c_1} \left(\frac{\cos\left( \frac{\alpha}{2} \right) }{\cos \left( \frac{\alpha_{0}}{2} \right) }\right)^{c_2} \sin \left( \alpha-\alpha_0 \right)~~,
        \end{equation}
\noindent where $\alpha_{0} $, $h$, $ c_1 $, and $ c_2 $ are free parameters for fitting polarization phase curves. We modify the original Lumme--Muinonen function so that $h$ corresponds to the polarimetric slope at $\alpha = \alpha_{0}$. We fit the observation data with Eq. (\ref{eq:trig}), employing the Markov chain Monte Carlo (MCMC) method in \texttt{emcee} (\citealt{Mackey2013}, version 3.0.2). 
We set boundary conditions of $0 < h < 1 $, $ 0 < c_1, c_2 < 10$, and $10\degr < \alpha_{0} < 40\degr$. The uncertainties of best-fit parameters are derived based on the 16th, 50th, and 84th percentiles of the samples in the marginalized distributions. The script to fit the polarization phase curve is available via the GitHub \footnote{ \url{https://github.com/Geemjy/Geem_2021_AA}}. 
The best-fit parameters with $\pm1\sigma$ uncertainties are summarized in Table \ref{T:fitting result}.
With the parameters we show the fitted profiles in Fig. \ref{Figure 2}. In the figure the polarization phase curves of Don Quixote and Hidalgo are similar to each other, but are different from QY1 in that they have polarimetric slopes steeper than QY1 (i.e., larger $h$ values). The inversion angle of Hidalgo is determined well to $\alpha_{0} = 18\fdg87^{+0.62}_{-0.84}$. On the contrary, the inversion angles of Don Quixote and QY1 were less precise due to the lack of data in the negative branch, preventing accurate determination (see Table \ref{T:albedo}), and yet the fitting process would work well because these $\alpha_{0}$ values are typical of general asteroids (i.e., $\sim 20\degr$).

\begin{table*}
\centering
\caption{Fitting result of polarization phase curves}
\label{T:fitting result}
\begin{tabular}{lccccccc}
\hline\hline
Target Name& Filter & ${h}$ & $\alpha_{0}$ & $c_1$ & $c_2$ & $\alpha_\mathrm{max}$&$P_\mathrm{max}$ \\
  &   & (\% / \mbox{deg}) & (\mbox{deg}) &    &   &(\mbox{deg})&(\%) \\
\hline
    Hidalgo &$R_\mathrm{C}$& $0.253^{+0.047}_{-0.056}$ & $18.87^{+0.62}_{-0.84}$ &$0.892^{+0.418}_{-0.351}$ & $5.429^{+3.178}_{-3.570}$  & $\ldots$    & $\ldots$   \\ 
    Don Quixote &$R_\mathrm{C}$& $0.353^{+0.223}_{-0.198}$ & $22.79^{+2.51}_{-3.70}$ & $1.017^{+0.871}_{-0.696}$ &$5.394^{+3.153}_{-3.535}$ &$\ldots$     &  $\ldots$  \\ 
    QY1& $V$ & $0.129^{+0.042}_{-0.050}$ &$24.05^{+4.24}_{-4.80}$& $0.323^{+0.402}_{-0.233}$ & $0.377^{+0.456}_{-0.267}$  & $98.83^{+5.91}_{-5.28}$ & $8.08^{+3.19}_{-3.11}$\\
    QY1 & $R_\mathrm{C}$& $0.127^{+0.043}_{-0.052}$& $24.87^{+4.03}_{-4.68}$ &$0.341^{+0.444}_{-0.248}$ &$0.347^{+0.406}_{-0.250}$ &$99.79^{+5.77}_{-5.26}$&$8.14^{+3.41}_{-3.55}$\\
\hline \\
\end{tabular}

\end{table*}

\begin{figure*}
                \centering
                \includegraphics[width=15cm]{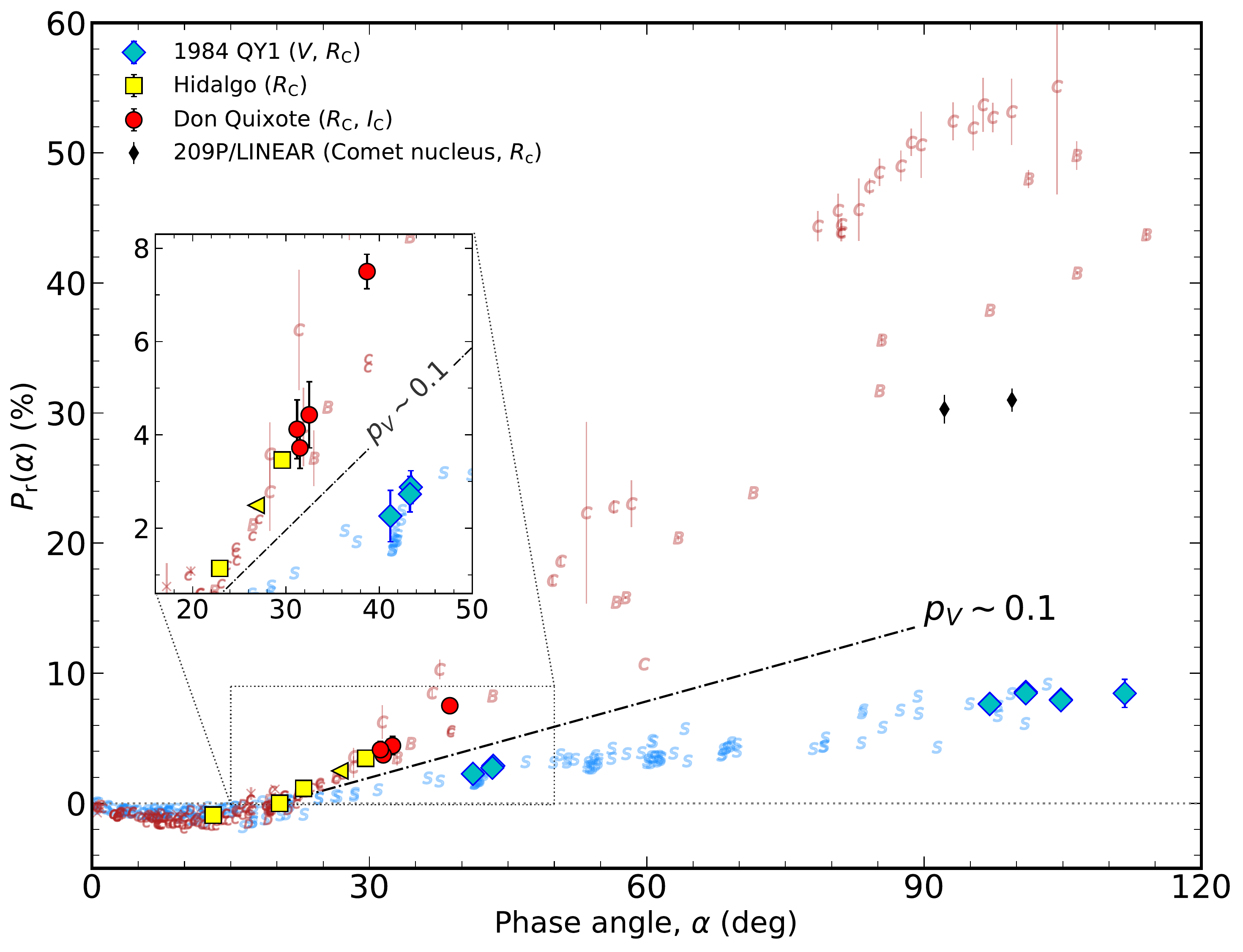}
                \caption{Comparison of $P_\mathrm{r}(\alpha)$ profiles of QY1, Don Quixote, and Hidalgo with asteroids and a bare comet nucleus, 209P/LINEAR,  \citep{2015ApJ...814..156K}. Each letter (C, B, and S) indicates the taxonomic types. The $P_\mathrm{r}$ value of Hidalgo in \citet{Fornasier2006} is plotted as the triangle marker. We plot the diagnostic line (dash-dotted line) corresponding to $p_\mathrm{V} = 0.1$. Objects with $p_\mathrm{V} \le 0.1$ are be located on the upper side of the line and vice versa. }
                \label{Figure 3}
        \end{figure*}

%Comparing with other asteroids in terms of polarimetry
Figure \ref{Figure 3} compares the polarization phase curves of three ACOs with a comet nucleus and other types of asteroids. We utilize the $P_\mathrm{r}$ data of a bare comet nucleus \citep[209P/LINEAR,][]{2015ApJ...814..156K} and C-, D-, B-, and S-type asteroids \citep{GIL2014,GIL2017,Ishiguro17,Ito2018,kuroda2021,Lupishko2014,Shinnka2018}.

In Fig. \ref{Figure 3}, the distribution of different objects depends mostly on their albedos. Lower albedo objects are distributed on the upper side, while higher albedo objects are distributed on the lower side. We draw the borderline corresponding to $p_\mathrm{V} = 0.1$ (dash-dotted line) in Fig. \ref{Figure 3}. The borderline is the straight line whose slope is obtained by putting $p_\mathrm{V} = 0.1$ in Eq. \ref{eq:slope} and starts from $\alpha = 20\degr$ (i.e., typical $\alpha_{0}$ of asteroids, \citealt{2017Icar..284...30B}). It becomes clear that objects with $p_\mathrm{V} < 0.1$ are located above the borderline. The $P_\mathrm{r}$ of Don Quixote, Hidalgo, and the comet nucleus are located on the upper side, indicating that they have consistent albedo values ($p_\mathrm{V} < 0.1$).  On the other hand, the polarimetric profile of QY1 is similar to that of S-type asteroids as it is below the line for $p_\mathrm{V} = 0.1$.

\begin{table*}
\centering
\caption{ Results of $\alpha_{0}$, $ h $, and $p_\mathrm{V}$}
\label{T:albedo}
\begin{tabular}{lccccc}
\hline\hline
Target Name& $\alpha_{0}$ & ${h}$ & ${\text{$p_\mathrm{V}$}}$ & References \\
  & (\mbox{deg}) & (\% /\mbox{deg} ) & & \\
\hline
    Hidalgo & ${18.87^{+0.62}_{-0.84}}^{a}$ &${0.253^{+0.047}_{-0.056}}^{a}$& ${0.050^{+0.017}_{-0.009}}^{b}$ &  $\ldots$    \\
    Don Quixote  & ${22.79^{+2.51}_{-3.70}}^{a}$ & ${0.353^{+0.223}_{-0.198}}^{a}$ &${0.035^{+0.049}_{-0.014}}^{b}$&   $\ldots$   \\
    QY1&${24.87^{+4.03}_{-4.68}}^{c}$& ${0.129^{+0.042}_{-0.050}}^{c}$  & ${0.153^{+0.107}_{-0.042}}^{b}$&  $\ldots$    \\
    \hline
  S-type asteroids&${20.7 \pm 0.2}^{c}$&$0.110 \pm 0.005 ^{c}$ &$0.21 \pm 0.08 $& 1, 2\\
  D-type asteroids&$18.2 \pm 0.3^{c}$& $0.341 \pm 0.109^{c}$&$0.09 \pm 0.05$& 1, 2 \\
  C-type asteroids&$19.4 \pm 0.1^{c}$&$0.387 \pm 0.037^{c}$&$0.07 \pm 0.04$& 1, 2 \\
  Comet nuclei&  $\ldots$  &  $\ldots$  &$0.02\sim 0.06$& 3 \\
\hline \\
\end{tabular}

\tablefoot{\\
\tablefoottext{a}{Derived by the observation in $R_\mathrm{C}$ band.}
\tablefoottext{b}{These $p_\mathrm{V}$ values were derived from $h$ and color.}
\tablefoottext{c}{Derived by the observation in $V$ band.}}
\tablebib{(1)~\citet{2017Icar..284...30B}; (2) \citet{2013ApJ...762...56U}; (3) \citet{Lamy2008}.
}
\end{table*}

\subsection{Derivation of geometric albedo}
\label{sec:reulst_albedo}

Since the polarimetric slopes depend on their albedo, as seen in Sect. \ref{Pr-alpha pro}, we derive the geometric albedos ($p_\mathrm{V}$) of ACOs from their slope $h$. It is well known that the slope $h$ has a good correlation with $p_{\mathrm{V}}$. This  was first noted by \citet{1967AnWiD..27..109W} and \citet{1967JGR....72.3105K}. The correlation is expressed with the  empirical equation
    \begin{equation}  
        \label{eq:slope}
        \log_{10} \left( p_\mathrm{V} \right) = C_1 \log_{10} \left( h \right) + C_2 ~~,
        \end{equation}
\noindent  where $ C_1 $ and $C_2 $ are constants. They have been determined by several research groups \citep{2012ApJ...749..104M, 2015MNRAS.451.3473C, Lupishko2018}. Here we use $ C_1 = -1.016 \pm 0.010 $ and $ C_2= -1.719 \pm 0.012 $ from \citet{Lupishko2018}, which uses the most comprehensive data sets obtained by infrared space telescopes and occultations. In addition, although $C_1$ and $C_2$ are obtained in $V$ band, we practically assume that the slope $h$ is dominantly controlled by the geometric albedo regardless of wavelength \citep{Umow1905} to apply $C_1$ and $C_2$ to our data in $R_\mathrm{C}$ band. Substituting the slope $h$ values in Eq. (\ref{eq:slope}), we computed the geometric albedo values; and  the results are summarized in Table \ref{T:albedo}. The uncertainty of the albedo is calculated based on the uncertainties of $h$, $C_1$, and $C_2$ in Eq. (\ref{eq:slope}). For comparison, we provide the average values or typical range of asteroids and comet nuclei in Table \ref{T:albedo}. 

In the case of Don Quixote and Hidalgo, only geometric albedos in the $R_\mathrm{C}$ band ($p_\mathrm{R_\mathrm{C}}$) are derived from our polarimetry. The results are $p_\mathrm{R_\mathrm{C}} = {0.055^{+0.077}_{-0.023}}$ for Don Quixote and $p_\mathrm{R_\mathrm{C}} =  {0.078^{+0.032}_{-0.015}}$ for Hidalgo. Since the geometric albedo is defined in the $V$ band,  we should convert them (i.e., $p_\mathrm{R_\mathrm{C}}$) to $V$-band albedos (i.e., $p_\mathrm{V}$) using their color indices ($V-R$). The applied $V-R$ values are summarized in Appendix \ref{App:h-VR}. The corresponding $p_\mathrm{V}$ values are summarized in Table \ref{T:albedo}. These $p_\mathrm{V}$ values of Don Quixote and Hidalgo are in the range of the typical $p_\mathrm{V}$ of comet nuclei and C- and D-type asteroids \citep{Lamy2008,2013ApJ...762...56U}. In contrast, $p_\mathrm{V}$ of QY1 is in the range of typical $p_\mathrm{V}$ of S-type asteroids \citep{2013ApJ...762...56U}.

We note that slope $h$ of Don Quixote and QY1 are derived by extrapolation to the range of $\alpha < 20\degr$ where no data is available. Because Eq. (\ref{eq:trig}) used for the fittings is the empirical function \citep{Lumme1993, 2005A&A...432.1081P}, polarimetric parameters derived by extrapolation is uncertain. However, we confirm that, while their $\alpha_{0} > 15\degr$, Don Quixote always shows the slope $h$ and the albedo (i.e., $p_\mathrm{V} < 0.1$) compatible with those of D-type asteroids (the optical analog of comet nuclei), whereas QY1 indicates these values are comparable with those of  S-type asteroids.

\subsection{Slope $h$ and the color Index $V-R$}
\label{sec:h-VR}
\begin{figure*}
                \centering
                \includegraphics[width=15cm]{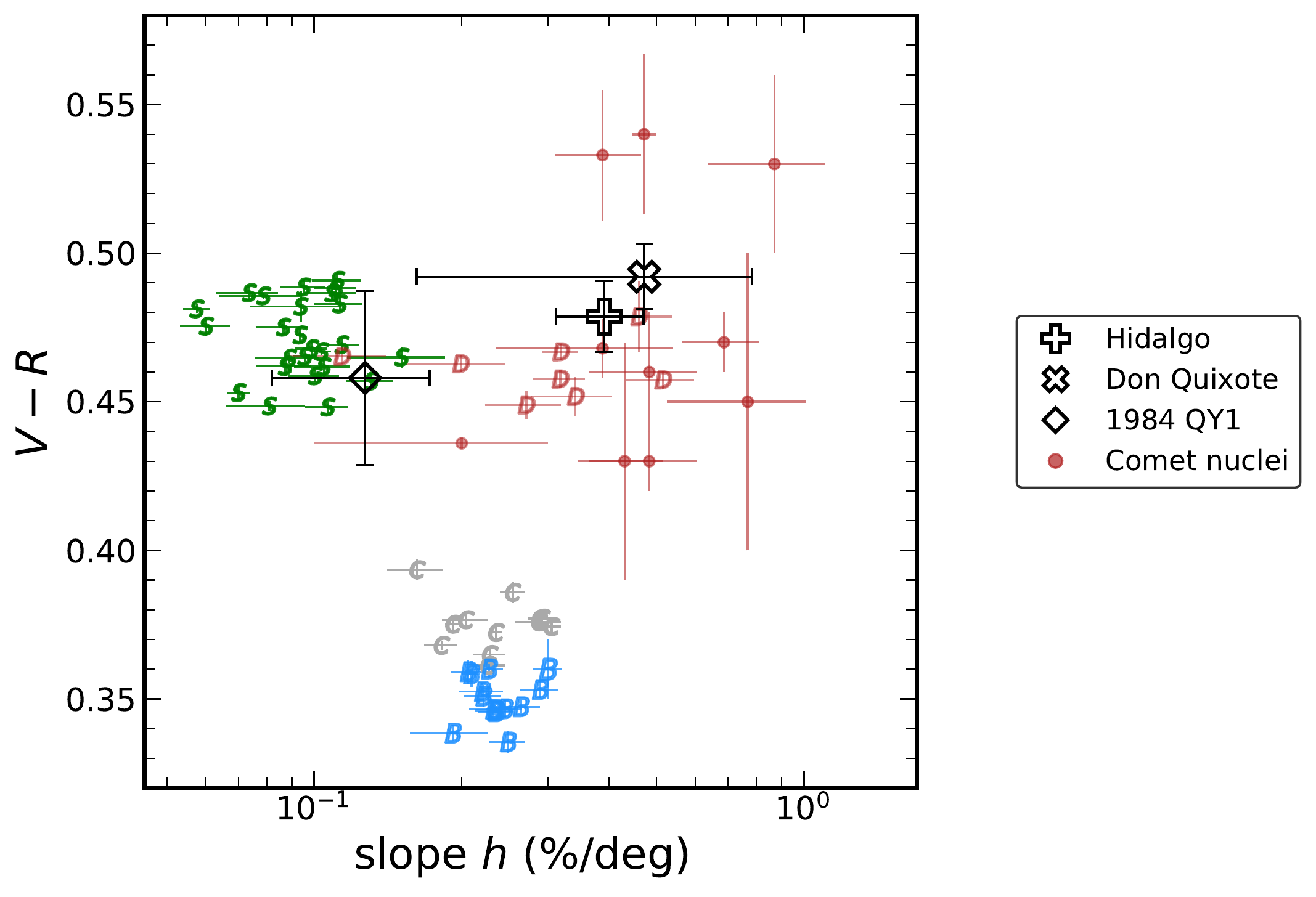}
                \caption{ Polarimetric slope $h$ and $V-R$ plot of Hidalgo, Don Quixote, and QY1 with   different types of asteroids and comet nuclei. Each letter (C, B, S, and D) indicates the taxonomic asteroid type. We label (7968) Elst-Pizarro as a B-type asteroid \citep{Licandro2011}. More details and references are given in Sect. \ref{sec:h-VR} and Appendix \ref{App:h-VR}.}
                \label{fig:VR}
        \end{figure*}

%VR-slope h diagram
Although the polarimetric slope $h$ is a useful proxy of albedo, it is insufficient to distinguish possible dormant comets from C-complex asteroids (C-, F-, and B-types) because comet nuclei (including D-type) and C-complex asteroids have similar albedo values. Therefore, we utilize the color index $V-R$ together with the slope $h$. We compare the slopes $h$ and $V-R$ of ACOs with those of other asteroids and comet nuclei (Fig. \ref{fig:VR}). We convert the slope $h$ of Don Quixote and Hidalgo in the $R_\mathrm{C}$ band to the $V$ band using their $V-R$ color indices (Appendix \ref{App:h-VR}).
In Fig. \ref{fig:VR}, objects are divided into three major groups: S-type asteroids; C-, F-, and B-type asteroids; and comet nuclei. Because comet nuclei have optical properties (colors and albedos) similar to D-type asteroids, they overlap with each other. Don Quixote and Hidalgo are clearly distinguished from C-type asteroids and are located in a  region similar to  comet nuclei and D-type asteroids.
Meanwhile, QY1 is compatible with S-type asteroids.

\section{Discussion}
\label{sec:discussion}

In this study we attempt to extract dormant comets from the ACO list. We conduct a polarimetric pilot survey for three ACOs to test the potentiality and found that two are likely dormant comets, while another is an S-type asteroid. Here we describe the characteristics of these three ACOs in the following subsections.

\subsection{(3552) Don Quixote}
Don Quixote should be in the class of comets of outer Solar System origin that contain volatile components such as H$_2$O and CO$_2$ ices. This object was discovered in 1983 as an asteroid despite the comet-like orbit \citep[$T_\mathrm{J} < 3$,][]{dormant_Asteroid3}.
It has a very elongated orbit with perihelion and aphelion distances of 1.24 $\mbox{au}$ and 7.28 $\mbox{au}$, respectively. The diameter and geometric albedo are estimated as $D = 18.4_{-0.4}^{+0.3}$ $\mbox{km}$ and $p_\mathrm{V}=0.03^{+0.02}_{-0.01}$ from thermal infrared data taken with the {\it Spitzer Space Telescope}, the NASA Infrared Telescope Facility, and the {\it Wide-field Infrared Survey Explorer} ({\it WISE}) \citep{Mommert_2014}. It is classified as a D-type asteroid \citep{Tholen1984, Bus2002,2003PASP..115..362R}. Recent telescopic observations at optical and infrared wavelengths confirm that Don Quixote has exhibited weak comet-like activity at heliocentric distances within 3 $\mbox{au}$ \citep{Mommert_2014, Mommert_2020,Kokhirova2021}. The activities were not episodic but recurrent, as observed at different perihelion passages in 2009 and 2017--2018. Moreover, a coma and a tail show the excess signal associated with CO$_2$ molecules in the {\it Spitzer Space Telescope} observation \citep{Mommert_2014}. For these reasons, there is no doubt that Don Quixote is a volatile-bearing cometary object of outer Solar System origin.

Our polarimetric observation was conducted starting on July 24, 2018, ten days after the cessation of activity was confirmed by \citet{Kokhirova2021}. We thus measure the polarization degree of the bare nucleus. Without using a space infrared telescope or a large telescope with a mid-infrared instrument, we derive the albedo of $p_\mathrm{V}=0.035^{+0.049}_{-0.014}$, which is consistent with the result from the {\it Spitzer Space Telescope}  \citep{Mommert_2014}. The comet-like optical properties are seen in the polarimetric  slope--color plot (Fig. \ref{fig:VR}). The polarimetry of Don Quixote thus becomes a benchmark for demonstrating the validity of dormant comet extraction using the polarimetric slope--color plot.

\subsection{(944) Hidalgo}
\label{sec:hidalgo}

Since its discovery in 1920 Hidalgo has never exhibited comet-like activity. Nevertheless, it is  suspected to be a dormant comet for the reasons described here. This object has a very elongated orbit with perihelion and aphelion distances of 1.95 $\mbox{au}$ and 9.53 $\mbox{au}$. Thus, this object not only intersects Jupiter's orbit (the semimajor axis $a = 5.20$ $\mbox{au}$),  but also reaches Saturn's orbit ($a = 9.55$ $\mbox{au}$) around its aphelion.
The diameter and geometric albedo are estimated as $D = 61.4 \pm 12.7$ $\mbox{km}$ and $p_\mathrm{V} = 0.028 \pm 0.006$ using {\it WISE} data and $D = 52.45 \pm 3.60$ $\mbox{km}$ \citep{Licandro2016} and $p_\mathrm{V} = 0.042 \pm 0.007$ using {\it AKARI} data \citep{Usui2011}. The albedo value we derived via polarimetry is $p_\mathrm{V} =0.050^{+0.017}_{-0.009}$. These result are in the albedo range of comet nuclei. Additionally, Hidalgo has a spectrum of D-type asteroids \citep{Tholen1984, Bus2002,2003PASP..115..362R}. All the results support the idea that Hidalgo is a strong candidate for a dormant comet.

As shown in Table \ref{table:2} and Fig. \ref{Figure 2}, Hidalgo is observed around $\alpha \sim 20\degr$, making it possible to derive the polarimetric inversion angle ($\alpha_{0}$). From the fitting we derive $\alpha_{0} ={18\fdg87^{+0.62}_{-0.84}}$. The derived $\alpha_{0}$ of Hidalgo is slightly smaller than for the majority of asteroids, but consistent with the typical $\alpha_{0}$ value of D-type asteroids (i.e., $\alpha_{0} = 18\fdg2 \pm 0\fdg3 $, \citealt{2017Icar..284...30B}), strengthening the result that it is  D-type. Meanwhile, there are two reports regarding the  $\alpha_{0}$ of  objects showing comet-like activity  derived without  gas or dust contamination: 2P/Encke, which indicates $\alpha_{0} \sim 13 \degr$ in the $R$ band \citep{2008A&A...489.1337B}, and (7968) Elst-Pizarro, which indicates $\alpha_{0} = 17\fdg6 \pm 2\fdg1$ in the $R$ band and $\alpha_{0} = 17\fdg0 \pm 1\fdg6$ in the $V$ band \citep{Bagnulo2010}. These $\alpha_{0}$ values are smaller than typical asteroids and are nearer to F-type asteroids \citep[$\alpha_{0} \sim 15\degr$,][]{2005Icar..178..213B,2016MNRAS.455.2091C,2017Icar..284...30B}.

\citet{Bagnulo2010} further note that three F-like asteroids, (4015) Wilson-Harrington (C- or F-type; \citealt{Tholen1984}), (3200) Phaethon (B- or F-type; \citealt{1989aste.conf..298T, Licandro2007}), and (155140) 2005 UD (B- or F-type; \citealt{2007A&A...466.1153K}), have evidence of dust emissions, and they  point out the association between small $\alpha_{0}$ asteroids and dust-ejecting objects. Although there are only two report (2P/Encke and (7968) Elst-Pizarro) that indicated the small $\alpha_{0}$, it is interesting to study the small $\alpha_{0}$ objects from the viewpoint of dust-ejecting objects. It is also a recent discovery that OSIRIS-REx witnesses dust ejection from (101955) Bennu \citep{2019Sci...366.3544L}. From   polarimetry it is reported that the asteroid has a small $\alpha_{0}$ \citep[i.e., $\alpha_{0}=17\fdg88\pm0\fdg40$,][]{Cellino2018}.

In this paper, however,  we consider that the application of $\alpha_{0}$ may not be a decisive factor to distinguish comets (including dust-emitting objects) from asteroids. Because the purpose of this study is to discriminate icy cometary objects of outer Solar System origin from asteroidal objects, we should regard the small $\alpha_{0}$ objects (7968) Elst-Pizarro (a Themis family member, \citealt{2004AJ....127.2997H}) and (101955) Bennu (an asteroid possibly originating from the Polana-Eulalia family complex, \citealt{2015Icar..247..191B}) as asteroids rather than comets. We note that our designations of comets and asteroids in this paper (Appendix \ref{app:Usage of term}) do not contradict the idea of \citet{Bagnulo2010}. As described in \citet{2005Icar..178..213B}, highly reflective particles with a size comparable to  the optical wavelength may affect the small $\alpha_0$ for 2P/Encke and (7968) Elst-Pizarro. As the number of $\alpha_{0}$ measurements for dust-ejecting objects increases in the future, it is expected that there may be a finding regarding the surface state of objects with small $\alpha_{0}$.

Thus, it is very likely that Hidalgo is a dormant comet,  even if it does not have a small $\alpha_{0}$.

\subsection{(331471) 1984 QY1}
\label{sec:qy1}
We conclude that QY1 is most likely an asteroid because of its high albedo ($p_\mathrm{V} =0.153^{+0.107}_{-0.042}$). The $ P_\mathrm{max} $ value ($8.14^{+3.41}_{-3.55}$ $\rm{\%}$ in the $R_\mathrm{C}$ band) is significantly lower than that of the 209P/LINEAR nucleus. Recent SMASSII observation data indicate that QY1 is an S$_\mathrm{q}$-type or Q-type asteroid when using the Bus-Demeo classification tool. The spectrum displays absorptions of approximately 0.9 $\mu$m and 1.9 $\mu$m, typical of these types of asteroids \citep{2009PDSS..114.....D, 2003PASP..115..362R}. Thus, these observations (including our polarimetry) indicate that QY1 is an S-complex asteroid.

\begin{figure}
                \resizebox{\hsize}{!}{\includegraphics{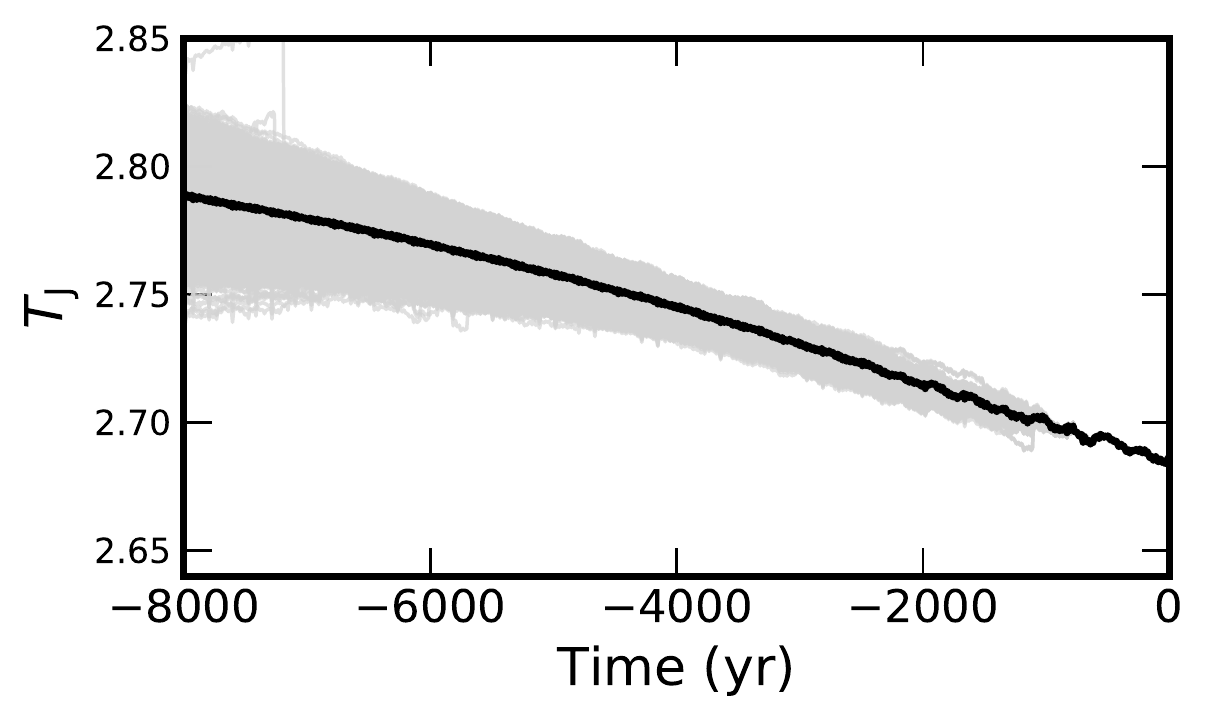}}
                \caption{ Time evolution of the Tisserand parameter ($T_\mathrm{J}$) with respect to Jupiter of QY1. $T_\mathrm{J}$ decreases by $ 0.1 $ over $ \sim $ 8\,000 years, which supports that QY1 would be transported from the main-belt region. Each gray line represents different clones whose current orbital elements follow a Gaussian distribution around the average values within their standard deviations (Table \ref{T:orbital}). The black line represents the results for a particle with the average orbital elements at Epoch 2459396.5 (2021-07-01.0).}
                \label{Fig:TJ}
        \end{figure}

The polarimetry of QY1 provides a rare opportunity for deriving the surface particle size. Taking advantage of our observations at large phase angles, we derive the particle size on QY1. It is known that $P_\mathrm{max}$ depends on the geometric albedo \citep{Umow1905} and the particle size \citep{1986MNRAS.218...75G}. 
\citet{1992Icar...99..468S} derived a formula to estimate the particle size $d$ (in $\mu$m) from $P_\mathrm{max}$ and a sort of albedo:

\begin{equation}
        \label{eq:Umow}
        d = 0.03 \exp\left( 2.9\left( \log_{10} \left( 100~A \right) + 0.845\log_{10} \left(10~P_\mathrm{max}\right)  \right)     \right) ~~,
        \end{equation}

\noindent where $A$ denotes an albedo at $\alpha=5\degr$. Applying the intensity ratio ($I(0\fdg3)/I(5\degr) = 1.44\pm0.04$ for S-type asteroids; \citealt{2000Icar..147...94B}), we obtain $A=0.11\pm0.03$ for QY1. Substituting $A$ in Eq. (\ref{eq:Umow}), we obtain an estimate of the particle diameter, $d\approx70$ $\mu$m. The size is slightly larger than the  S-type asteroid (4179) Toutatis \citep[$\lesssim50$--80 $ \mu$m,][]{1997PASJ...49L..31I,2019JKAS...52...71B}, but smaller than the near-Sun Q-type asteroid (1566) Icarus \citep[100--130 $\mu\mbox{m}$,][]{Ishiguro17}.
We note that Eq. (\ref{eq:Umow})  should be applied to asteroids carefully since a formula is established on the lunar samples. Even so, we use  Eq. (\ref{eq:Umow}), which is the same method as in previous studies for comparison.

Lastly, we consider the paradoxical problem that the S-type asteroid QY1 has a comet-like orbit. According to a dynamical study of near-Earth objects \citep{2002Icar..156..399B}, QY1 has a 96.1 $\rm{\%}$ probability of Jupiter-family comet origin, which was one of the highest-potential dormant comet candidates in the list. Since then, its orbital elements have been updated thanks to the accumulation of astrometric observations, yet QY1 has $ T_\mathrm{J}=2.68$, which is significantly smaller than the criterion of $ T_\mathrm{J}=3$. From the revised semimajor axis of $ a=2.497$ $\mbox{au}$ we note that QY1 is in a 3:1 mean motion resonance (MMR) with Jupiter (i.e., $ a_{3:1} = 2.50 \pm 0.03 $ $\mbox{au}$). As pointed out in \citet{2014ApJ...789..151K} and \citet{2014Icar..234...66T}, $ T_\mathrm{J}$ must be treated carefully  when considering origins. Main-belt asteroids in resonance should experience increasing orbital eccentricity to be transported into the near-Earth region \citep[e.g.,][]{2002aste.book..409M}. QY1 has likely been injected into the current comet-like orbit by means of the 3:1 MMR with Jupiter. To confirm this hypothesis, we conduct a backward dynamical simulation of QY1 considering the gravity of eight planets and the Sun (Fig. \ref{Fig:TJ}). We employ the \texttt{Mercury 6} integrator for the simulation \citep{1999imda.coll..449C}. We generate 200 clones with the orbital elements of QY1 within the 1$\sigma$ range at the current epoch, considering their orbit covariances (quoted from the JPL Small-Body Database Browser site\footnote{\url{https://ssd.jpl.nasa.gov/}}). The applied orbital elements and their uncertainties are summarized in Table \ref{T:orbital}. We integrate these parameters to 8\,000 years in the past with a time step of 8 days, considering the gravitational forces of the Sun and eight planets but ignoring the Yarkovsky force. As shown in Fig. \ref{Fig:TJ}, although the $ T_\mathrm{J} $ values disperse before $-$1\,500 years, there is a general trend that $ T_\mathrm{J} $ values continue decreasing over time ($ \Delta T_\mathrm{J} \approx -0.1 $ for 8\,000 years). Based on this dynamic integration and our polarimetric results, QY1 would be an object of main-belt origin rather than outer Solar System origin. This result is consistent with the fact that S-complex asteroids are dominant in the 3:1 MMR (45 $\rm{\%}$, \citealt{2014PASJ...66...51K}). For confirmation, we examine the possible source regions using the updated source region probability models \citep{2012ApJ...749L..39G,Granvik2018} and find that QY1 has a high possibility ($ \approx$60 $\rm{\%}$) of  main-belt origin in 3:1 MMR and a very low probability ($ \approx $1--3 $\rm{\%}$) of  Jupiter-family comet origin using the updated orbital elements in Table \ref{T:orbital}.

        \begin{table*}
                \caption{Orbital elements of QY1 at Epoch 2459396.5 (2021-07-01.0)}
                \label{T:orbital}
                \centering
                \begin{tabular}{c c c c c c}
                        \hline\hline
                        $a^a$ & $e^b$ & $i^c$ & $g^d$ & $n^e$ & $M^f$\\
                        (\mbox{au})&&(\mbox{deg})&(\mbox{deg})&(\mbox{deg})&(\mbox{deg})\\
                        \hline
                        $2.500163$  & $0.893873$  & $14.282613 $
                        & $ 337.182719 $ & $142.265114 $ 
                        & $113.120460$  \\
                        $ \pm 8.660 \times 10^{-9}$ & $ \pm 2.001 \times 10^{-8}$ & $\pm 5.270 \times 10^{-6}$ & $\pm 2.192 \times 10^{-5}$ & $\pm 2.240 \times 10^{-5}$ & $\pm 2.865 \times 10^{-6}$\\
                        \hline
                \end{tabular}
                \tablefoot{                                  
                        \tablefoottext{a}{Semimajor axis in \mbox{au},}
                        \tablefoottext{b}{Eccentricity, }
                        \tablefoottext{c}{Inclination in \mbox{deg}, }
                        \tablefoottext{d}{Mean argument of perihelion in \mbox{deg},}
                        \tablefoottext{e}{Longitude of the ascending node in \mbox{deg},}
                        \tablefoottext{f}{Mean anomaly in \mbox{deg.}}\\
                        We obtained these elements from the web-based JPL Small-Body Database Browser (https://ssd.jpl.nasa.gov/sbdb.cgi).
                }
        \end{table*}    
\subsection{Potentiality of polarimetry for ACO research}
Finally, we describe the effectiveness of polarimetric observations in ACO research. The discrimination of dormant comets from the ACO population is challenging because both have a point-source appearance. The geometric albedos and reflectance spectra (or color indices) have been considered for discrimination. Comet nuclei have featureless and reddish colors due to irradiated organic materials on their surface \citep{Meech2004, Licandro2011}, while asteroids have a wide variety of colors \citep{1989aste.conf..298T,Binzel2004}. Similarly, comet nuclei have low albedos \citep[typically $p_\mathrm{V} = 0.02$--$0.06 $,][]{2002EM&P...89..117C,2004come.book..223L}, while asteroids have a wide range of albedos ($ p_\mathrm{V} =  0.02$--$0.60$, \citealt{Usui2011}). D-type asteroids have  optical properties similar to those of  comet nuclei so they are indistinguishable by spectroscopic or photometric observation \citep{2009Icar..202..160D, Licandro2011}.

To date, geometric albedos of asteroids are derived mostly by radiometry. In the radiometric method, albedo values are derived from the combination of absolute magnitudes and sizes obtained via a thermal model with observation data. Accordingly, albedo values derived by radiometry have inherent uncertainties related to the applied thermal model and optical magnitudes. In addition, the use of mid-infrared observation facilities (e.g., space telescopes, such as {\it AKARI}, {\it IRAS}, and {\it Spitzer}, or ground-based telescopes with a mid-IR camera, such as SUBARU/COMICS) is becoming difficult (as of July 2021). On the other hand, because polarimetric instruments are less expensive than these infrared instruments, they are installed in a relatively large number of small and intermediate-sized telescopes. Using such instruments, albedo estimations, which were conventionally performed in infrared space telescopes or large telescope facilities on the ground, are possible. Since albedos can be obtained directly from polarimetric parameters using the empirical equation, there is no other information \citep{1967AnWiD..27..109W, 1967JGR....72.3105K, CELLINO20122552, 2015MNRAS.451.3473C}. Additionally, the constant parameters for deriving albedos from polarimetry  continue to be updated and are becoming more reliable \citep{2012ApJ...749..104M, 2015MNRAS.451.3473C, Lupishko2018}. In addition, polarimetry with a polarizing beam splitter is highly feasible even under variable conditions, canceling out variable weather conditions to produce reliable results.

In the future it is expected that a large number of ACOs will be discovered by large systematic surveys, especially by the Vera C. Rubin Observatory (previously known as the Large Synoptic Survey Telescope, {\it LSST}; \citealt{Vera2021}). Early follow-up polarimetric observations with a small or intermediate-sized telescope with a polarimetric instrument are expected to provide an overview of the dormant comet population lurking in the inner Solar System. Our work will be helpful in that we demonstrated the effectiveness and potentiality of polarimetry by conducting this ACO pilot survey for three objects.

\section{Summary}
        We conducted a polarimetric pilot survey for three ACOs (Don Quixote, Hidalgo, and QY1). These three ACOs have a $T_\mathrm{J}$ value significantly smaller than three, and they were recognized as highly possible dormant comet candidates \citep{Hartmann1987,2002Icar..156..399B}. We obtain the polarization phase curve to conjecture their origins together with color information from previous studies. Our major findings are the following:
        \begin{enumerate}
                \item Don Quixote and Hidalgo show polarimetric and color profiles similar to those of comet nuclei and D-type asteroids. Their albedos derived by our polarimetric data are in the range of comet nuclei.
                \item Our result of Don Quixote is consistent with the fact that the object indicated recurrent comet-like activities around its perihelion passages. Hidalgo is also likely  a dormant comet.
                \item The polarimetric profile of QY1 was unexpected, showing a profile similar to S-type asteroids. We find  from the
dynamical simulation that QY1 was transported from the main belt via the 3:1 mean motion resonance with Jupiter.
                \item QY1 has $8.08^{+3.19}_{-3.11}$ $\rm{\%}$ in the $ V $ band and $8.14^{+3.41}_{-3.55}$ $\rm{\%}$ in the $R_\mathrm{C} $ band. From the $ P_\mathrm{max} $ values we obtain an estimate of the particle diameter on the surface of QY1 of  $d\approx70$ $\mu$m.
        \end{enumerate}

The remaining issue is the polarimetric inversion angle ($\alpha_0$). Hidalgo's $\alpha_0$ is in the range of D-type asteroids and the active asteroid (7968) Elst-Pizarro, but out of 2P/Encke's range. Further polarimetric observations of comet nuclei and ACOs around the inversion angle are required to determine the inconsistency.

\begin{acknowledgements}
        This work at Seoul National University was supported by the National Research Foundation of Korea (NRF), funded by the Korean government (MEST; No. 2018R1D1A1A09084105). The Pirka telescope is operated by the Graduate School of Science, Hokkaido University and is partially supported by the Optical \& Near-Infrared
Astronomy Inter-University Cooperation Program, MEXT, Japan. Part of the spectral data utilized in this publication was obtained and made available by the MIT-UH-IRTF Joint Campaign for NEO Reconnaissance. The IRTF is operated by the University of Hawaii under Cooperative Agreement no. NCC 5-538 with the National Aeronautics and Space Administration, Office of Space Science, Planetary Astronomy Program. The MIT component of this work is supported by NASA grant 09-NEOO009-0001 and by the National Science Foundation under Grants Nos. 0506716 and 0907766. Finally, we appreciate the anonymous reviewer for providing constructive and encouraging comments and suggestions.    \end{acknowledgements}

%%%%%%%%%%%%%%%%%%%%%%%%%%
%%%    APPENDIX     %%%%%%
%%%%%%%%%%%%%%%%%%%%%%%%%%

\begin{appendix}

\section{Usage of the terms comets and asteroids}
\label{app:Usage of term}
Here we describe the designations of comets and asteroids adopted throughout this paper. 

Comets and asteroids have been distinguished from several viewpoints (such as their appearance, composition, and orbital properties). If a small body of the Solar System indicates activity accompanied by a coma and a tail, it is conventionally regarded as a comet; otherwise, it is regarded as an asteroid. From their composition, asteroids consist mostly of refractory components with a small amount of volatiles (or without volatiles), while comets are rich in volatile components and refractory components. 

Comets and asteroids are also distinguished by their orbital properties.   Comets are thought to migrate from the outer Solar System (i.e., the Kuiper Belt or the Oort Cloud) via dynamic interactions with Jovian planets, and to exhibit comet-like activities when they receive extra solar radiation that causes ice sublimation to form comae and tails. Such objects from the outer Solar System intersect with Jovian planets and have  Tisserand parameter values  $ T_\mathrm{J}<3$. On the other hand, asteroids are dynamically disconnected from Jovian planets and have $ T_\mathrm{J}>3$ \citep{1997Icar..127...13L}.

These comet-asteroid discrimination methods do not always work, however. For example, 2P/Encke contains icy volatiles (e.g., H$_2$O and CO$_2$, \citealt{2013Icar..226..777R}) showing regular activity, but has the Tisserand parameter $ T_\mathrm{J}=3.03$ (i.e., an asteroidal orbit). It is considered that 2P/Encke has the current asteroidal orbit due to the nongravitational effect (acceleration by sublimation of ice) and the gravitational interaction with planets \citep{2006Icar..182..161L}. Furthermore, after discovering the so-called main-belt comets (comets in the main asteroidal belt with $ T_\mathrm{J}>3$), the distinction between comets and asteroids became ambiguous \citep{2006Sci...312..561H}. The recent discovery of very red asteroids (similar to Kuiper Belt objects) further complicates the designations \citep{Hasegawa_2021}.

We were motivated to study ACOs by pioneering research: \citet{Fernandez1997}, \citet{2008Icar..194..436D}, \citet{2008A&A...487.1195L}, and \citet{2014ApJ...789..151K}. ACOs are asteroids with comet-like orbits ($ T_\mathrm{J}<3$), and most ACOs are thought to be dormant (or low activity) comets that are at the last stage of the evolution of the bodies from the Kuiper Belt or the Oort Cloud, although there are some asteroids transported to the current orbits via mechanisms such as the Yarkovsky effect \citep{2002aste.book..409M, 2014ApJ...789..151K}. Since the purpose of this study is to distinguish comets from ACOs,  throughout this paper we refer to objects originating from or in the main belt as asteroids and objects from the Kuiper Belt or the Oort Cloud as comets.

                \section{Photometry of QY1}
                
We conducted these observations to detect a signature of comet-like activity.           
A series of photometric observations were made at three observatories: the Okayama Astrophysical Observatory (OAO), the Ishigakijima Astronomical Observatory (IAO), and the Observatoire de Haute-Provence (OHP). The detailed circumstances of the observations are given in Table \ref{photometry}.

The OAO is located atop Mt. Chikurinji, Okayama Prefecture, Japan (133\degr35\arcmin 36\arcsec E,  34\degr 34\arcmin33\arcsec N,  360 $\mbox{m}$). We performed  observations on three nights on UT May 2--5, 2016, using the Multicolor Imaging Telescopes for Survey and Monstrous Explosions (MITSuME) with three Alta U6 cameras ($1024 \times 1024$ $\mbox{pixels}$) attached to the 50 $\mbox{cm}$ telescope. The IAO is located on Ishigaki Island, Okinawa Prefecture, Japan (124\degr 08\arcmin 21\farcs4 E, 24\degr22\arcmin22\farcs3 N, 197 $\mbox{m}$). We observed the target asteroid for seven nights on UT May 26--June 12 2016. We used the 105 cm Murikabushi Cassegrain telescope and  MITSuME. The MITSuME at IAO was identical  to the system at OAO. The OHP is located in Alpes-de-Haute-Provence, Saint-Michel-l'Observatoire, France (5\degr42\arcmin48\arcsec E, 43\degr55\arcmin51\arcsec N, 650 $\mbox{m}$). We made observations on three nights on UT August 1--3 2016. We utilized a 120 cm telescope (focal length of 7.2 $\mbox{m}$) and an Andor Ikon L 936 camera ($2048 \times 2048$ $\mbox{pixels}$). The fields of view and pixel scales of these instruments are 12\farcm3 $\times$ 12\farcm3 (0\farcs72 pixel$^{-1}$) at IAO, 26\farcm0 $\times$ 26\farcm0 (1\farcs53 pixel$^{-1}$) at OAO, and 13\farcm1$ \times $13\farcm1 (0\farcs38 pixel$^{-1}$) at OHP.

We note that all of these images show a point-like target without showing a comet-like coma and tail. The brightness modulation by rotation was detected, as shown below. The observed raw magnitudes were converted into magnitudes viewed at heliocentric and geocentric distances of $r_\mathrm{h}=\Delta=1 ~\mbox{au}$ and a phase angle of $\alpha=70\degr$,

\begin{eqnarray}
                m_\mathrm{R} \left( 70\degr \right)=m_\mathrm{R} - 5~\log_{10} \left( r_\mathrm{h} \Delta \right)+ b \left(\alpha-70\degr \right)~~,
                \label{eq:appendix}
                \end{eqnarray}
                \noindent
                where the phase coefficient of $b=0.032$ was assumed, which is a predicted value for an object with $p_{\rm V}=0.178$ \citep[see the empirical equation on page 99,][]{2000Icar..147...94B}.

We made a plot of the light curve applying the rotation periods in \citet{warner2016}, where two rotational periods are suggested:  $ P_1 = 45.5 $ hours and $ P_2 = 36.6 $ hours (Fig. \ref{photometry}). $ P_1 $ is the best candidate of the main period. The uncertainty of the period of 0.5 hours is quoted. From our light curve  data, the modulation of $P_1$ is clearly seen, but the modulation of $P_2$ is not, probably because our amount of data may not be sufficient to find it.

\begin{figure*}
        \centering
\includegraphics[width=18cm]{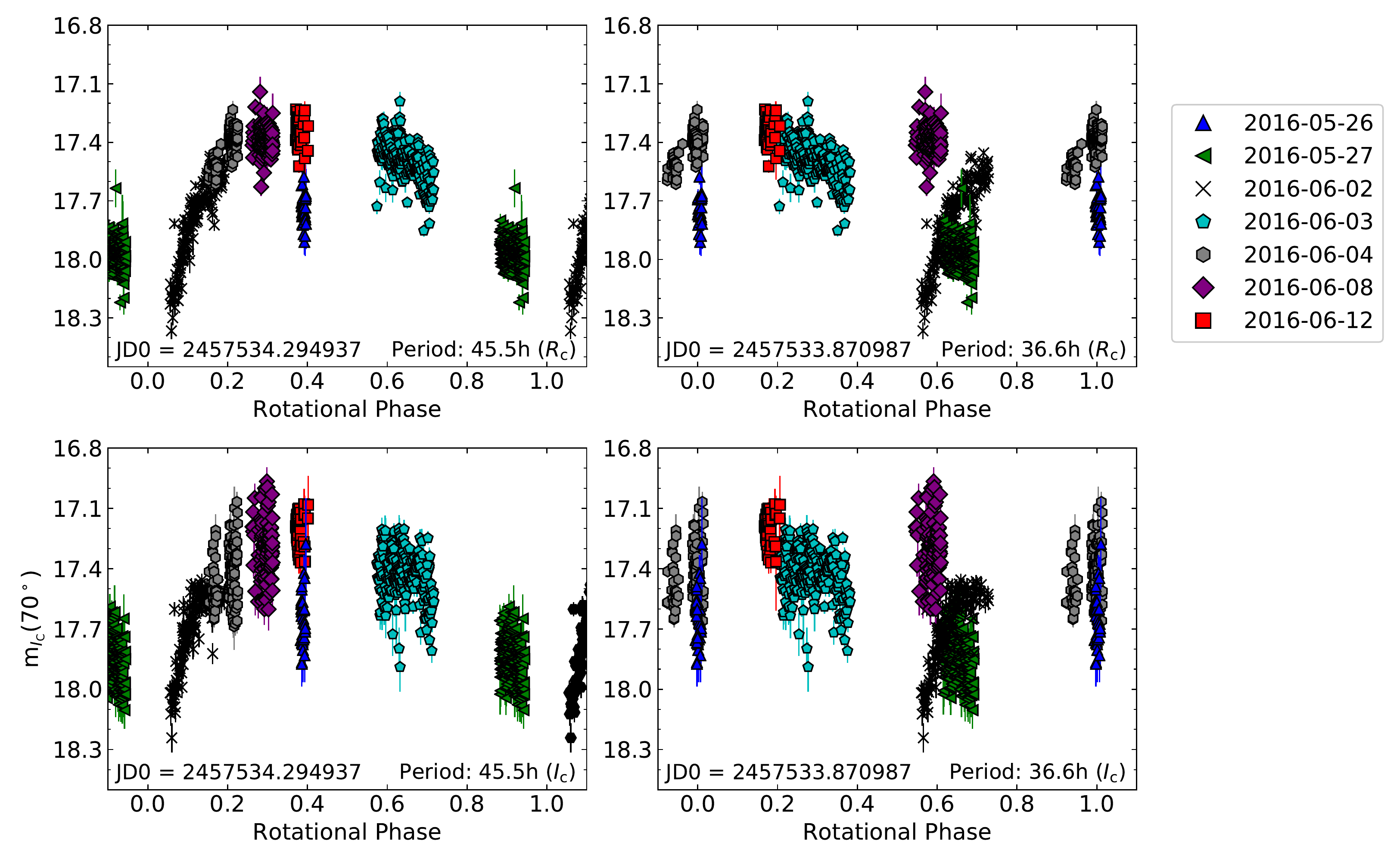}
\caption{Light
curves of $R_\mathrm{C}$ band (top) and $I_\mathrm{C} $ band (bottom)  produced assuming a rotational period of $P_1 = 45.5$ hours (left column) and $P_2 = 36.6$ hours (right column) \citep[see][]{warner2016}. The horizontal axes indicate the rotational phase, and the vertical axes indicate the magnitudes viewed from $r_\mathrm{h}=\Delta=1 ~\mbox{au}$ at $\alpha=70\degr$, assuming a phase slope parameter $b=0.032 $ $\mbox{mag}$/\mbox{deg}.}
\label{Fig:light}
\end{figure*}

The observations at OHP was conducted approximately two months after the observations at OAO and IAO. The accuracy of the rotational period (0.5 hours) is not sufficient to compare the OHP data with the others, so we do not plot the OHP data in Fig. \ref{photometry}. The magnitudes at OHP, $m_\mathrm{R}(70\degr)=17.20$--$17.65$ $\mbox{mag}$, are in the range of the maximum and minimum magnitudes in Fig. \ref{photometry}, so it is likely that the phase angle correction with $b$ works well for the OHP data. This  supports the validity of our estimate for the geometric albedo. The reduced magnitude of QY1 in Fig. \ref{Fig:light} is available at the CDS\footnote{\url{http://cdsweb.u-strasbg.fr/cgi-bin/qcat?J/A+A/XX}}.

\begin{table*}
                        \caption{Observation circumstance of photometric data}
                        \label{photometry}
                        \centering
                        \begin{tabular}{l c c c c c c c c }
                                \hline\hline
                                Date & UT& Telescope&Filter&Exptime$^a$  & $ N^b $ &$ r^c $&$ \Delta^d $ &$ \alpha^e $ \\
                                & & & & (\mbox{sec})& & (\mbox{au})& (\mbox{au}) &(\mbox{deg}) \\
                                \hline
                                2016-May-26 & 12:33--13:09 & IAO & $g'$, $ R_\mathrm{C} $, $ I_\mathrm{C} $ & 30 & 76 & 0.88 & 0.29 & 108.6 \\
                                2016-May-27 & 11:15--14:03 & IAO & $g'$, $ R_\mathrm{C} $, $ I_\mathrm{C} $  & 20 & 476 & 0.90 & 0.28 & 105.0 \\
                                2016-Jun-02 & 11:39--17:48 & OAO&$g'$, $ R_\mathrm{C} $, $ I_\mathrm{C} $  & 120 & 320 & 1.01 & 0.28 & 82.1 \\
                                2016-Jun-02 & 13:05--14:17 & IAO &$g'$, $ R_\mathrm{C} $, $ I_\mathrm{C} $ & 20 & 18 & 1.01 & 0.28 & 82.2 \\      
                                2016-Jun-03 & 11:14--17:43 & OAO&$g'$, $ R_\mathrm{C} $, $ I_\mathrm{C} $ & 120 & 350 & 1.03 & 0.29 & 78.5 \\
                                2016-Jun-03 & 11:47--17:16 & IAO & $g'$, $ R_\mathrm{C} $, $ I_\mathrm{C} $ & 20 & 384 & 1.03 & 0.29 & 78.5 \\
                                2016-Jun-04 & 13:34--16:58 & IAO &$g'$, $ R_\mathrm{C} $, $ I_\mathrm{C} $  & 20 & 222 & 1.05 & 0.29 & 75.1 \\
                                2016-Jun-05 & 11:47--16:20 & OAO&$g'$, $ R_\mathrm{C} $, $ I_\mathrm{C} $ & 120 & 164 & 1.07 & 0.30 & 72.0 \\            
                                2016-Jun-08 & 13:34--15:52 & IAO &$g'$, $ R_\mathrm{C} $, $ I_\mathrm{C} $ & 20 & 262 & 1.12 & 0.33 & 63.4 \\
                                2016-Jun-12 & 13:31--14:55 & IAO &$g'$, $ R_\mathrm{C} $, $ I_\mathrm{C} $  & 20 & 82 & 1.19 & 0.38 & 54.8 \\
                                2016-Aug-01 & 20:50 & OHP & $ R_\mathrm{C} $ & 300 & 1 & 1.90 & 1.39 & 31.3 \\
                                2016-Aug-02 & 20:23--20:29 & OHP &$ R_\mathrm{C} $  & 300 & 2 & 1.91 & 1.41 & 31.2 \\
                                2016-Aug-03 & 20:09--20:19 & OHP & $ R_\mathrm{C} $ & 300 & 3 & 1.92 & 1.43 & 31.0\\ 
                                \hline
                        \end{tabular}
                        \tablefoot{
                                The observation circumstance of light curve data taken at Okayama Astrophysical Observatory (OAO) and Ishigakijima Astrophysical Observatory (IAO). We used the web-based JPL Horizon system (http://ssd.jpl.nasa.gov/?horizons) to obtain these quantities.
                                \tablefoottext{a}{Exposure time in $\mbox{sec}$,}
                                \tablefoottext{b}{Number of data obtained,}
                                \tablefoottext{c}{Median heliocentric distance in \mbox{au},}
                                \tablefoottext{d}{Median geocentric distance in \mbox{au},}
                                \tablefoottext{e}{Median solar phase angle in \mbox{deg}.} 
                        }
                \end{table*}

        \section{Derivation of spectral gradients in Fig. \ref{fig:VR} }
        \label{App:h-VR}
        In Sect. \ref{sec:h-VR} we plotted the polarimetric slope $h$ and the color index $V-R$ for three ACOs, asteroids, and comet nuclei (Fig. \ref{fig:VR}). We used the $h$ values of asteroids provided in the catalog of the asteroid polarization curves \citep{GIL2017}. The applied slope $h$ values from the catalog are determined in the $V$ band. Because of the lack of $h$ data for D-type asteroids, we computed $h$ values from $p_\mathrm{V}$ values of 267, 1542, 2246, 2569, 2872, 3248, and 4744 by Eq. (\ref{eq:slope}). The albedos of these D-type asteroids are obtained from \citet{Usui2011}, \citet{Nugent2016}, and \citet{Tedesco2004}. If there is multiple albedo information in these catalogs, the averaged values are calculated and used for the plot. The $V-R$ of asteroids and ACOs are derived using the Small Main-Belt Asteroid Spectroscopic Survey (SMASS) data \citep{Bus2002,2003PASP..115..362R}. From SMASS spectra, we calculated the normalized spectral gradient ($S'$) defined as

\begin{equation}
        \label{eq:spectral gradient1}
        S' = \left(\frac{dS}{d\lambda}\right)\bigg/\:\overline{S}~~,
        \end{equation}

\noindent where $S$ is the $\lambda$-dependent reflectance, and $\overline{S}$ is the average $S$ in the wavelength range of $d\lambda$. Here the  $dS / d\lambda$ values were calculated by the linear fitting of SMASS spectra between 5\,500 $\angstrom$ and 6\,500 $\angstrom$. The derived $S'$ values were converted to $V-R$ values by using Eq. (2) in \citet{Jewitt2002}. We obtained $V-R = 0.49 \pm 0.01$ for Don Quixote and $V-R = 0.48 \pm 0.01$ for Hidalgo.

Because there is no optical spectrum for QY1, we derived it using our photometric data ($g'$-, $R_\mathrm{C}$-band). The value of 
    $S'$ of QY1 is derived as        
        
\begin{equation}
        \label{eq:spectral gradient of QY1}
        \log_{10}(S_\lambda) = \frac{m_\lambda - m_{\sun,\lambda}}{-2.5}~~
        \end{equation}
        \noindent and
        \begin{equation}
        \label{eq:spectral gradient of QY1 (2)}
        S' = \left(\frac{S_{R_\mathrm{C}} - S_{g'}}{\lambda_{R_\mathrm{C}} - \lambda_{g'} } \right)\bigg/\:\left(\frac{S_{R_\mathrm{C}}+ S_{g'}}{2}\right)~~,
                \end{equation}

\noindent where the subscript $\lambda$ denotes the effective wavelength of filters, and
$m_\lambda$ and $m_{\sun,\lambda}$ are the apparent magnitudes of the object and the Sun at wavelength $\lambda$, respectively. Here, we use $m_{\sun, R_\mathrm{C}} = -27.15$ $\mbox{mag}$ and $m_{\sun, \mathrm{g'}} = -26.34$ $\mbox{mag}$ \citep{Willmer_2018} and $\lambda_{R_\mathrm{C}}$ = 6\,480 $\angstrom$ and $\lambda_{g'}$ = 4\,710 $\angstrom$. With these parameters we derived the $V-R$ of QY1 as $V-R=0.46 \pm 0.03$.

Similarly to the D-type asteroids, the slope $h$ values of comet nuclei were computed from their albedos. The geometric albedos and $V-R$ (or $S'$) of comet nuclei were obtained from various sources \citep{Jewitt2002, Meech2004, Abell2005,Campins2006, Fernandez_2006, Lamy2008, Tubiana2008, Li2012}. For 2P/Encke and (7968) Elst-Pizarro, we referred to the slope $h$ values in \citet{2008A&A...489.1337B} and \citet{Bagnulo2010}.

\end{appendix}

\bibliographystyle{aa}
\bibliography{ref}

\end{document}